# Resurrection of Schrödinger Cat

Jae-Seung Lee[1], A. K. Khitrin[1]

[1]Department of Chemistry, Kent State University, Kent OH 44242-0001, USA

**Abstract**

Quantum decoherence is the major obstacle in using the potential of engineered quantum dynamics to revolutionize high-precision measurements, sensitive detection, or information processing. Here we experimentally demonstrate that quantum state of a system can be recovered after the state is destroyed by uncontrollable natural decoherence. Physical system is a cluster of seven dipolar-coupled nuclear spins of single-labeled $^{13}$C-benzene in liquid crystal. $^{13}$C spin plays a role of a device for measuring protons' "cat" state, a superposition of states with six spins up (*alive*) and six spins down (*dead*). Information about the state, stored in the $^{13}$C spin, is used to bring the protons' subsystem into the *alive* state, while the excess entropy produced by decoherence is transferred to the "measuring device", the $^{13}$C spin.

The most striking difference between quantum and classical systems is the ability of quantum objects to be in a superposition state. A system in a superposition of macroscopically distinct states (*alive* and *dead* states of the "Schrödinger cat") would demonstrate highly unusual behavior. Recent proposals suggest that the "cat" state can play an important role in quantum-enhanced measurements: high-precision spectroscopy[1,2], amplified detection[3,4], and quantum state measurement[5]. Decoherence resulting from interaction with environment is the major obstacle in designing practical devices. Here we experimentally demonstrate that quantum state of an object can be recovered after destruction by uncontrollable natural decoherence. Physical system is a cluster of seven dipolar-coupled nuclear spins of single-labeled $^{13}$C-benzene in liquid crystal. $^{13}$C spin plays a role of a device for measuring protons' "cat" state, a superposition of states with six spins up (*alive*) and six spins down (*dead*). Information about the state, stored in the $^{13}$C spin, is used to bring the protons' subsystem into the *alive* state, while the excess entropy produced by decoherence is transferred to the "measuring device", the $^{13}$C spin. Therefore, after the system's quantum state is irreversibly destroyed by decoherence, one part of the system can be brought into a desired pure state by using information stored in the other part.

Quantum state can be described by a wave function |Ψ> or, equivalently, by the corresponding density matrix ρ = |Ψ><Ψ|. This density matrix represents an ensemble of identical systems, each in the same quantum state |Ψ>. Such an ensemble is said to be in a pure state. When a system is in one of the two states: |u> (*alive*) or |d> (*dead*), in a basis where |u> and |d> are eigenstates, the density matrix has only one non-zero matrix element (Figs. 1a and 1b). For the superposition state $\frac{1}{\sqrt{2}}$ (|u>+|d>), the density matrix of

this pure state contains four non-zero elements (Fig. 1c). Two diagonal elements are the populations and the two off-diagonal elements describe a coherence between the two states. Interaction with environment can destroy the coherence. The result of this process, called decoherence, is the mixed state of an ensemble, shown in Fig. 1d. Individual systems are no longer in the same quantum state but can be found in one of the two states, |u> or |d>, with equal probability.

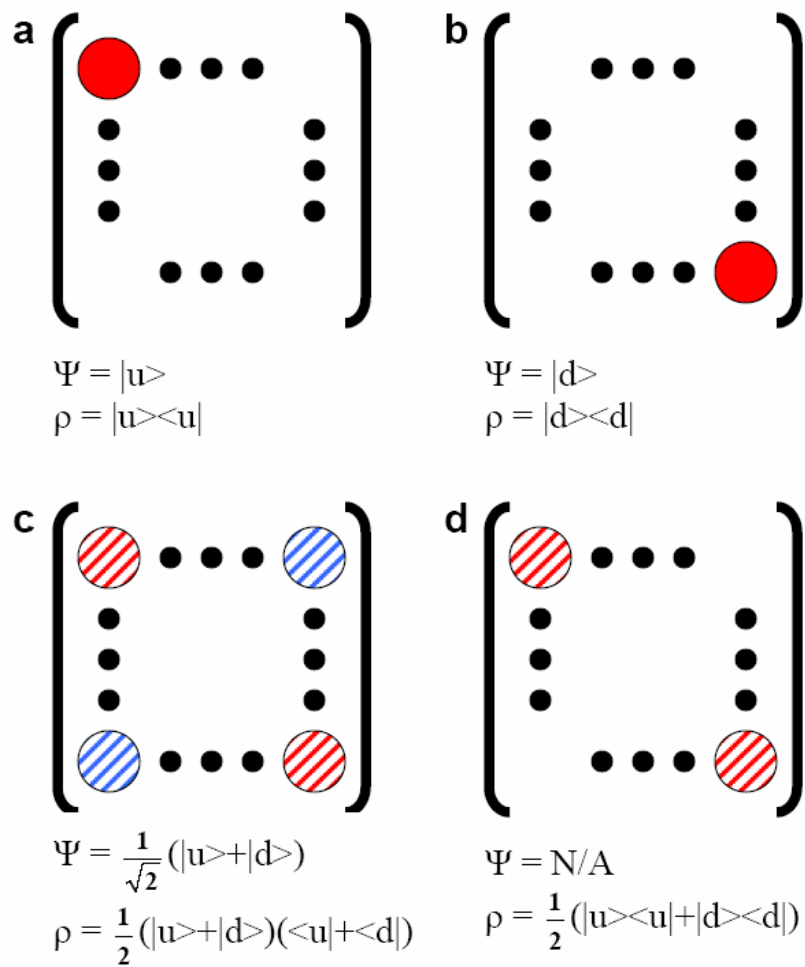

**Figure 1. The density matrices corresponding to four states. a,** the *alive* state |u>; **b,** the *dead* state |d>; **c,** the superposition state $\frac{1}{\sqrt{2}}$ (|u>+|d>); **d,** the mixture of the states |u> and |d>.

To be more specific, let us consider a system of $N$ coupled two-level systems (qubits). A cluster of spins ½ is a natural physical implementation of such system. As the two most distinct states, one can choose the ferromagnetic states $|u\rangle = |\uparrow\uparrow\ldots\uparrow\uparrow\rangle$ and $|d\rangle = |\downarrow\downarrow\ldots\downarrow\downarrow\rangle$ with all spins up and down, respectively. These two states have the maximum difference in polarization (magnetization). The off-diagonal part, $N$-quantum ($NQ$) coherence, of the cat state density matrix $\rho = \frac{1}{2}(|u\rangle+|d\rangle)(\langle u|+\langle d|)$ (Fig. 1c) has the operator form $|u\rangle\langle d| + |d\rangle\langle u| = S_1^+ S_2^+ \ldots S_N^+ + S_1^- S_2^- \ldots S_N^-$, where $S_i^+$ and $S_i^-$ are the raising and lowering operators for individual spins. When individual spins are rotated by angles $\varphi_i$ around their quantization axes, the two product operator terms in the $NQ$ coherence acquire the phases $\pm\sum_i \varphi_i$. Therefore, only the sum of the phases is important and, as an example, rotation of all spins by $\varphi$ and rotation of a single spin by $N\varphi$ produce the same result. This unique feature of the cat state opens a way to interesting applications. First, $N$ times faster rotation of the phase of the cat state, compared to that of an individual spin, can be used in high-precision measurements of phases and frequencies[1]. On the other hand, local rotation of a single spin can produce a global change of the phase for the entire system, which can be then converted into a change of macroscopic observables. This creates a base for amplified quantum detection[3,4] and state measurement[5].

The negative side of these useful properties is high sensitivity of the cat state to a phase noise. When interaction with environment produces uncorrelated rotations of individual spins by $\Delta\varphi_i$, the change in the phase of the $NQ$ coherence, $\Delta\varphi$, can be estimated as $\langle\Delta\varphi^2\rangle \approx N \langle\Delta\varphi_i^2\rangle$. This explains why even weak interaction with environment may cause fast decoherence in macroscopic systems. Random phases acquired by the *NQ* coherences of individual systems average out the off-diagonal part of the cat state density matrix (Fig. 1c) and convert it to the diagonal mixed state (Fig. 1d). Decoherence can be viewed as a loss of information about the phase of the *NQ* coherence of an individual system. This also means a loss of reversibility because, in order to dynamically convert a state into some target state, one needs to know the exact starting state. A degree of irreversibility is quantified by the entropy, which changes from zero for the pure cat state to $k_B \ln 2$ for the mixed state in Fig. 1d.

Different states of the same system have different sensitivity to noise. Some of them are resistant to relaxation and decoherence[6-9]. The cat state is the most fragile in terms of the decoherence rate, but the damage produced to this state by decoherence is relatively small. In quantum mechanics, density matrix gives the most complete description of a system. This means that no physical measurement can distinguish between the mixed state $\rho = \frac{1}{2}(|u\rangle\langle u|+|d\rangle\langle d|)$, resulting from averaging out the *NQ* coherences, and the mixture of systems in one of the two states, $|u\rangle$ or $|d\rangle$. Therefore, the only missing information is which of the two states has been chosen by a system. This information can be stored in only one additional qubit.

Let us consider a combined system of *N*+1 qubits. The symbols $|\uparrow\rangle$ and $|\downarrow\rangle$ will denote the two states of the additional qubit. It will be called a control qubit because its state is used to control a change of the state of the *N*-qubit system. We will start with the system in some superposition of the states $|u\rangle$ and $|d\rangle$ and the control qubit in its ground state $|\uparrow\rangle$:

$$|\Psi\rangle_{in} = |\uparrow\rangle (a |u\rangle + b |d\rangle), \qquad |a|^2 + |b|^2 = 1. \tag{1}$$

By using interaction between the control qubit and the system, one can design a reversible unitary evolution[5], which converts the state (1) into the state

$$|\Psi\rangle_{out} = a\,|\uparrow\rangle\,|u\rangle + b\,|\downarrow\rangle\,|d\rangle. \qquad (2)$$

Decoherence eliminates the $(N+1)Q$ coherence in this state and produces the mixed state with the density matrix

$$\rho_{mix} = aa^*\,|\uparrow\rangle\langle\uparrow| \otimes |u\rangle\langle u| + bb^*\,|\downarrow\rangle\langle\downarrow| \otimes |d\rangle\langle d|. \qquad (3)$$

In this mixture of two states, information about the state of the $N$-qubit system (is it $|u\rangle$ or $|d\rangle$) is stored in the state of the control qubit. It is possible to use this information for producing a dynamic evolution, conditioned by the state of the control qubit, which will bring the $N$-qubit system to a desired pure state, as an example, the *alive* state $|u\rangle$.

Alternatively, the state in Eq. (3) can be viewed as a result of projective measurement, where the control qubit is used as a "measuring device". The initial superposition state of the $N$-qubit system, $a\,|u\rangle + b\,|d\rangle$, is collapsed into one of the two definite states $|u\rangle$ or $|d\rangle$, while the remaining uncertainty is compensated by the measurement result, stored in the state of the control qubit.

For experimental implementation of this scheme we have chosen a cluster of seven dipolar-coupled nuclear spins in $^{13}$C-labeled benzene molecule oriented by a liquid-crystalline matrix. The sample is 5% of single-labeled $^{13}$C-benzene (Aldrich) dissolved in liquid crystal MLC-6815 (EMD Chemical). Each benzene molecule contains seven nuclear spins, one $^{13}$C and six protons, coupled by residual dipole-dipole interactions. All intermolecular spin-spin interactions are averaged out by fast molecular motions. Therefore, the system is an ensemble of non-interacting spin clusters, where each benzene molecule contains seven dipolar-coupled nuclear spins. The experiment has been performed with a Varian Unity/Inova 500 MHz nuclear magnetic resonance (NMR) spectrometer at room temperature (25°C).

At present, clusters of nuclear spins explored with NMR provide the largest and the most complex systems with individually addressable quantum states. With liquid-state NMR, coherent manipulations have been demonstrated for systems of up to seven spins[10,11].

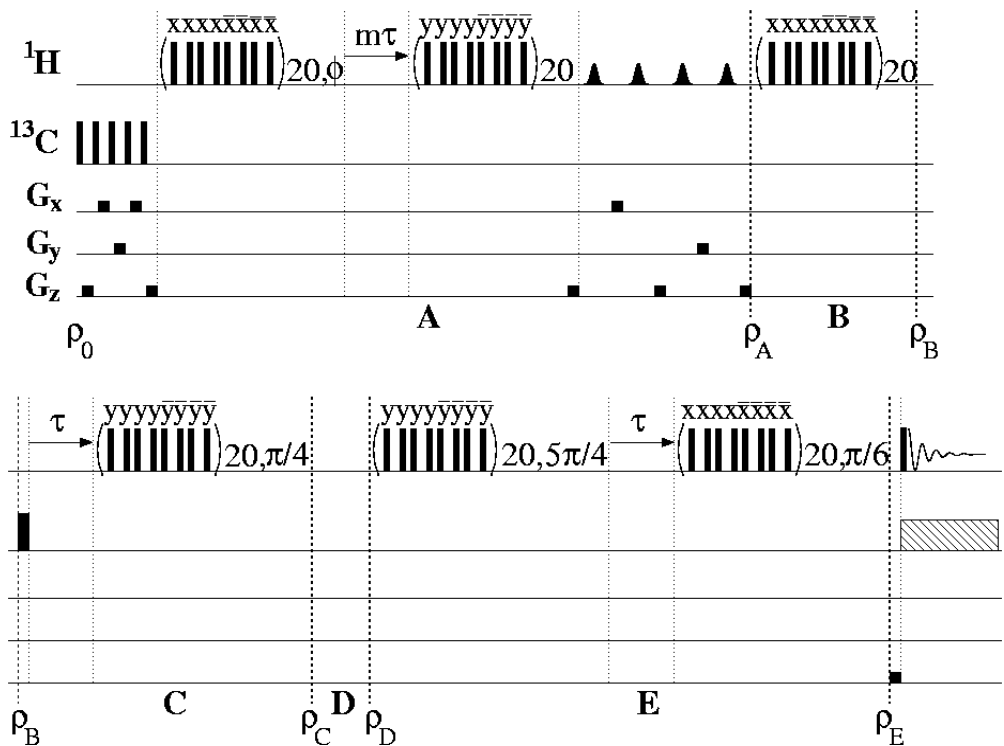

**Figure 2. NMR pulse sequence.** (Step A) the "cat" (six proton spins) and "measuring device" (carbon spin) are initialized in the pseudopure state; (step B) the cat is prepared in a superposition state; (step C) the cat is entangled with the measuring device; (step D) decoherence brings the whole system into a mixed state; (step E) the "controlled-not" operation, conditioned on the carbon spin state, creates the target state of the proton subsystem.

At thermal equilibrium, nuclear spins are in a highly mixed state, which means that an individual system can be, with some probability, in any of possible quantum states. The most convenient way to address individual states is by using so-called pseudopure states. The idea of state initialization by creating a pseudopure state was originally proposed for NMR-based quantum computing[12-14]. In a pseudopure state, populations of all but one state are made equal. As a result, the spin density matrix is a sum of a maximally mixed background, which is proportional to the unity matrix, and a deviation part, which is proportional to a density matrix of a pure state. Since the unity matrix does not contribute to observables and is not changed by unitary transformations, behavior of a system in pseudopure state is exactly the same as it would be at zero spin temperature. With liquid-state NMR, pseudopure states of a seven-spin system have been demonstrated[10], and the off-diagonal part of the cat state has been prepared as a benchmark in another seven-spin system[11]. For dipolar-coupled spins, we have recently created pseudopure states[15], including the cat state[16], for a system of twelve nuclear spins of fully $^{13}$C-labeled benzene in liquid crystal.

Creation of the pseodopure state $|\uparrow\rangle|u\rangle$ for our seven-spin system is the first part (step A) of the experimental scheme in Fig. 2. It includes two multiple-quantum evolution periods implemented with multi-pulse sequences[17], filtering the highest-order coherence, evolution caused by dipole-dipole interaction between the carbon and protons' spins, and partial saturation with selective *sinc*-pulses. The details of the technique are described elsewhere[3,5,15]. $^1$H linear-response spectra of the four pseudopure states are shown in Fig. 3. The states $|\uparrow\rangle|d\rangle$, $|\downarrow\rangle|u\rangle$, and $|\downarrow\rangle|d\rangle$ are obtained from the state $|\uparrow\rangle|u\rangle$ by applying "hard" 180° carbon and proton pulses. The left column in Fig. 3 displays the spectra with decoupled $^{13}$C spin. All the spectra coincide precisely with the theoretical spectra calculated for these states[5]. Each of the decoupled spectra (left column) consists of a single peak. The $^{13}$C-coupled spectra for the states $|\uparrow\rangle|u\rangle$ (Fig. 3b') and $|\downarrow\rangle|d\rangle$ (Fig. 3e') have two peaks, and there are one large and two smaller peaks in the spectra for the states $|\uparrow\rangle|d\rangle$ (Fig. 3c') and $|\downarrow\rangle|u\rangle$ (Fig. 3d'). For comparison, the thermal equilibrium spectra are shown in Figs. 3a and 3a'.

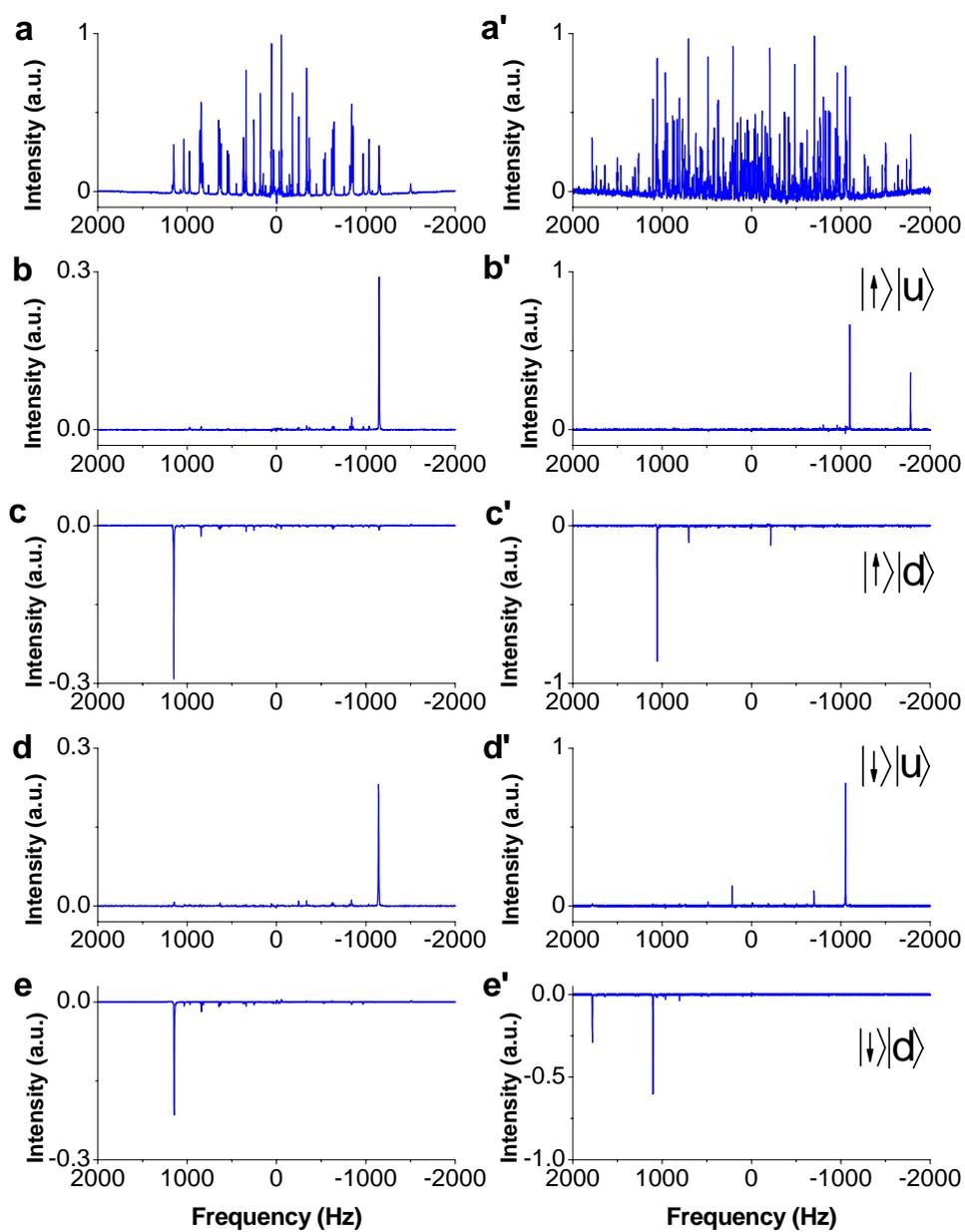

**Figure 3**. **Linear-response $^1$H NMR spectra at thermal equilibrium (a, a') and for four pseudopure states (b – e and b' – e').** The left column shows $^{13}$C-decoupled spectra.

Step B of the experiment (Fig. 2) creates the six-spin cat state for the proton subsystem and, therefore, the total state becomes that of Eq. (1) (at $a = b = 1/\sqrt{2}$), which is the starting point of our "resurrection" scheme.

The next step C (Fig.2) is designed to convert the protons' six-spin cat state of Eq. (1) into the seven-spin cat state of the entire system, described by Eq. (2). This step includes 90° carbon pulse, carbon-proton evolution delay, and protons multiple-quantum evolution period. The $^{13}$C-decoupled and $^{13}$C-coupled spectra of this seven-spin cat state are shown in Figs. 5a and 5a', respectively.

The variable-length step D is a delay when decoherence takes place and converts the pure state of Eq. (2) into the mixed state of Eq. (3). Before proceeding to the next step, we measured the decay times of the off-diagonal, i.e. the seven-quantum (7Q) coherence, and the diagonal elements of the seven-spin cat state density matrix. The results are shown in Fig. 4. Since the 7Q coherence is not directly observable, we followed step D by an additional step, which is a time-reversal of step C (not shown in Fig. 2). After that, the reading pulse produced the spectra with the intensities of the peaks proportional to the intensity of the 7Q coherence. The decay of the diagonal elements of the cat state has been measured by simply applying a reading pulse after step D. Different symbols in

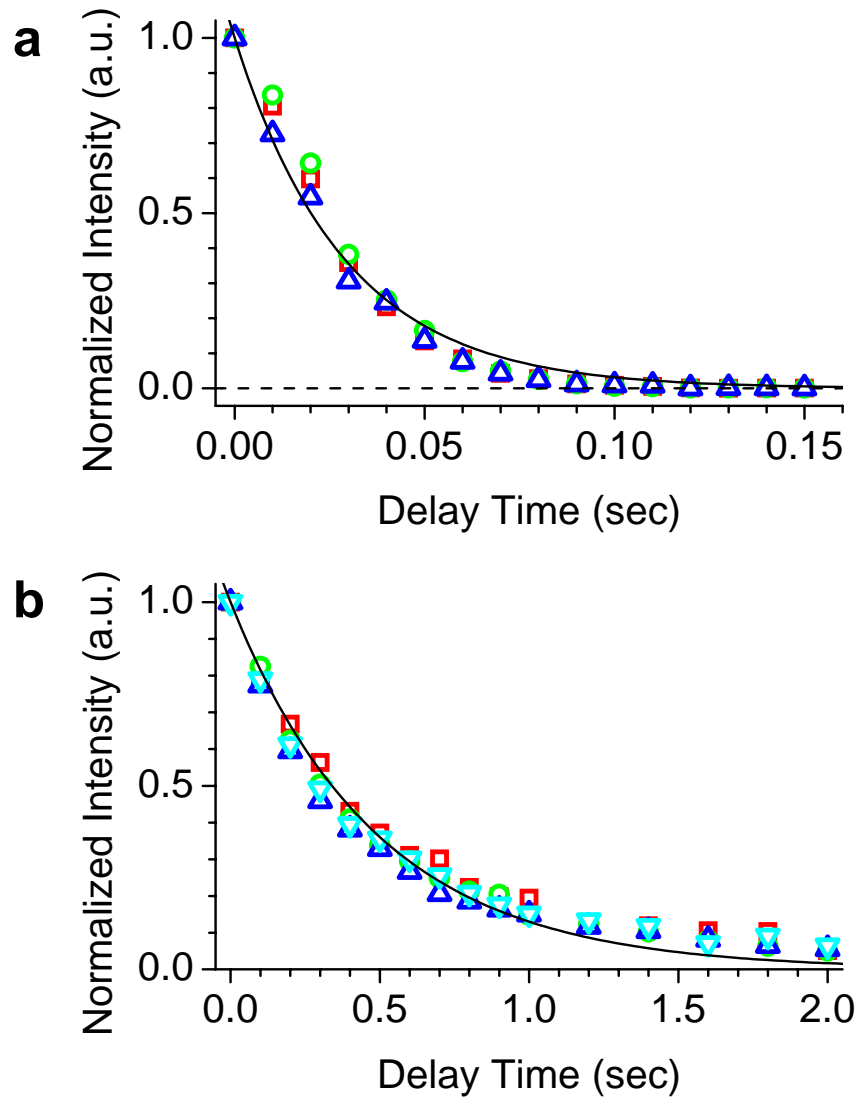

**Figure 4. Lifetime of the seven-spin cat state. a,** Decay of the 7Q coherence. The solid line is the function exp(-t/$\tau$) with $\tau$ = 0.029 sec. **b,** Decay of the diagonal states. The solid line is the function exp(-t/$\tau$) with $\tau$ = 0.49 sec.

Figs. 4a and 4b show the results obtained by using intensities of different peaks in the spectra. Decays of both the off-diagonal (Fig. 4a) and the diagonal (Fig. 4b) elements of the cat state are well described by single-exponential curves with the average lifetimes of 29 ms and 490 ms, respectively. Please note that the time scales in Figs. 4a and 4b are different. The lifetime of the 7Q coherence, or the decoherence time of the cat state, is much shorter than the relaxation time of its diagonal elements. This difference between the decoherence and the relaxation times is supposed to increase with the system size. For our seven-spin cluster, the difference is already significant and allows clean implementation of the proposed scheme.

The last step E of the experiment (Fig. 2) implements the "controlled-not" operation: when the state of the control qubit is |↑>, it does not change the state of the proton subsystem; when the state of the qubit is |↓>, it flips the protons state |d> into the state |u> and vice versa. As a result, a mixture of the two states |↑>|u> and |↓>|d> after step D is converted into a new mixture of the states |↑>|u> and |↓>|u>. In both of these states the proton subsystem is in the target *alive* state |u>. This is supported by the $^{13}$C-decoupled spectra in Figs. 5c and 5d for 100 ms and 200 ms delays in step D. The spectra are the same as the spectrum of the pseudopure state |u> in Figs. 3b and 3d. One can see that even after 200 ms delay time, which is much longer than the seven-spin cat state decoherence time (29 ms) or the protons six-spin cat state decoherence time (42 ms), the target state |u> is well recovered. At the same time, the carbon spin is found in the mixture of the two states |↑> and |↓>, which can be clearly seen in the $^{13}$C-coupled spectra in Figs. 5c' and 5d'. These spectra are the sums of the spectra for the two pseudopure states |↑>|u> and |↓>|u> in Figs. 3b' and 3d'. Therefore, the excess entropy $k_B$ln2 produced by decoherence is transferred to the $^{13}$C spin.

One can notice that quality of the spectra in Fig. 5 is somewhat lower than that for the pseudopure states in Fig. 3. It is a consequence of complex dynamics and very long total sequence of 982 radio-frequency and gradient pulses. However, even with unavoidable imperfections, the experimental results present a convincing proof of the principle. They show that after a time much longer than the cat state decoherence time, the information recorded in the "measuring device", the $^{13}$C spin, is sufficient to bring the protons subsystem into a desired target state. With resetting the control qubit or supplying a new one, it is possible to repeat the whole process. However, the scheme does not protect the system from relaxation, an energy-exchange process, which decreases intensities of the spectra at increasing delay time (Figs. 5c and 5d).

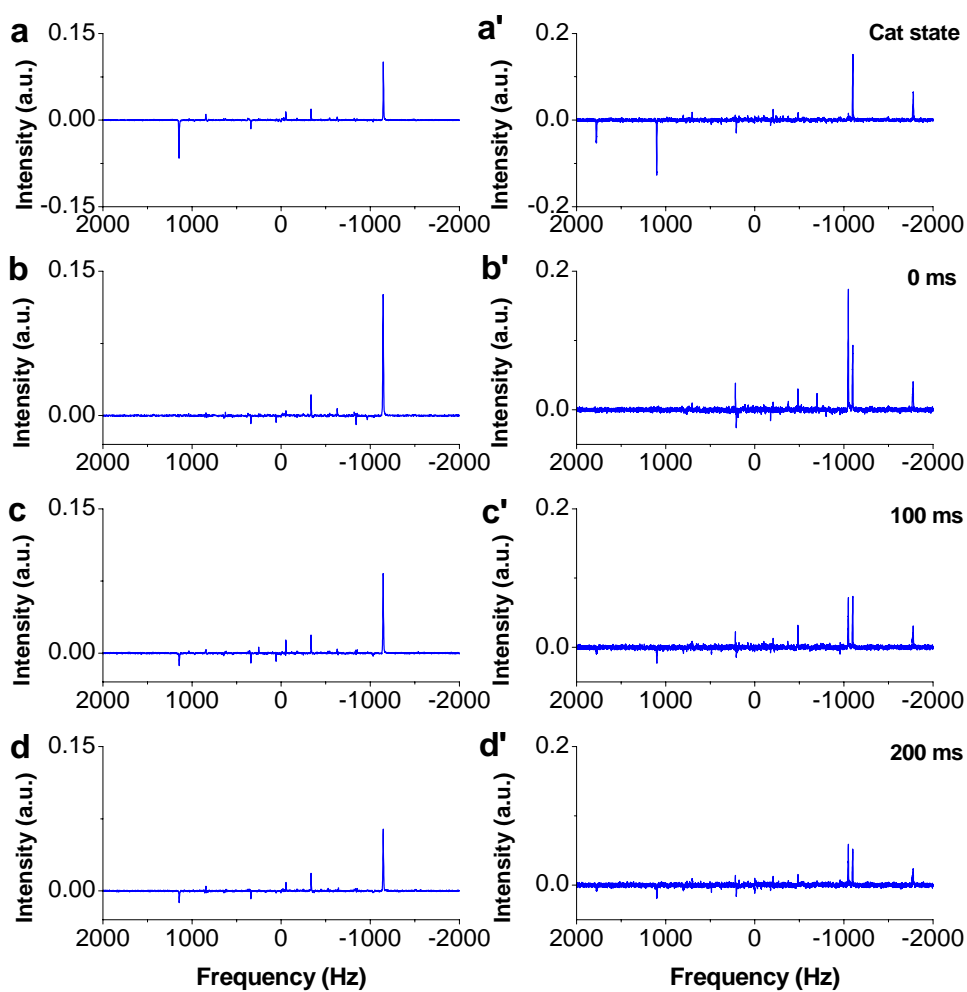

**Figure 5. Linear-response ¹H NMR spectra for the seven-spin cat state (a, a'), and the restored six-spin *alive* state |u⟩.** The delay times are (b, b') 0 ms, (c, c') 100 ms, and (d, d') 200 ms. The left column shows $^{13}$C-decoupled spectra.

Our resurrection scheme resembles an active quantum error-correcting algorithm[18]. In both cases, additional resources are used to bring the system back to the target pure state after decoherence or errors. Simple quantum error-correcting schemes have been experimentally realized for small quantum systems[19-21]. While these experiments demonstrated protection from artificial errors and engineered decoherence effects, our experiment implements a state recovery after uncontrollable natural decoherence.

Decoherence is the major obstacle in designing practical devices which will utilize the advantages of quantum dynamics in detection, measurement, or information processing. The experiment performed in this work demonstrates that destructive effect of decoherence can be minimized by using a scheme where information stored in one part of a system is used to restore quantum state of another part.